\documentclass[aps,prd,nofootinbib,twocolumn,superscriptaddress,preprintnumbers]{revtex4}

\usepackage{subfigure}
\usepackage{amssymb}
\usepackage{amsmath}
\usepackage{epsfig}
\usepackage{hyperref}

\usepackage{graphicx,array}
\usepackage{color}
\usepackage{latexsym}
\usepackage{amsthm}
\usepackage{amsmath}

\usepackage{subfigure}
\makeatletter
\def\simgt{\mathrel{\lower2.5pt\vbox{\lineskip=0pt\baselineskip=0pt
           \hbox{$>$}\hbox{$\sim$}}}}
\def\simlt{\mathrel{\lower2.5pt\vbox{\lineskip=0pt\baselineskip=0pt
           \hbox{$<$}\hbox{$\sim$}}}}
\makeatother

\newcommand{\be}{\begin{equation}}
\newcommand{\ee}{\end{equation}}
\newcommand{\bea}{\begin{eqnarray}}
\newcommand{\eea}{\end{eqnarray}}

\newcommand{\Fig}[1]{Fig.~\ref{#1}}
\newcommand{\Figs}[2]{Figs.~\ref{#1} and \ref{#2}}
\newcommand{\Eq}[1]{Eq.~(\ref{#1})}

\newcommand{\Sec}[1]{Sec.~\ref{#1}}

\newcommand{\OO}{\mathcal{O}}

\newcommand{\gxmin}{g_{X,\text{min}}}
\newcommand{\gxmax}{g_{X,\text{max}}}
\newcommand{\sBR}{\sigma \text{BR}}

\begin{document}

\title{Higgs Mass from D-Terms: a Litmus Test}

\author{Clifford Cheung} \author{Hannes L.~Roberts}
\affiliation{Berkeley Center for Theoretical Physics,
  University of California, Berkeley, CA 94720, USA}
\affiliation{Theoretical Physics Group,
  Lawrence Berkeley National Laboratory, Berkeley, CA 94720, USA}

\begin{abstract}

We explore supersymmetric theories in which the Higgs mass is boosted by the non-decoupling D-terms of an extended $U(1)_X$ gauge symmetry, defined here to be a general linear combination of hypercharge, baryon number, and lepton number.  Crucially, the gauge coupling, $g_X$, is bounded from below to accommodate the Higgs mass, while the quarks and leptons are required by gauge invariance to carry non-zero charge under $U(1)_X$. This induces an irreducible rate, $\sigma$BR, for $pp \rightarrow X \rightarrow \ell\ell$ relevant to existing and future resonance searches, and gives rise to higher dimension operators that are stringently constrained by precision electroweak measurements. Combined, these bounds define a maximally allowed region in the space of observables, ($\sigma$BR, $m_X$), outside of which is excluded by naturalness and experimental limits.  If natural supersymmetry utilizes non-decoupling D-terms, then the associated $X$ boson can only be observed within this window, providing a model independent `litmus test'  for this broad class of scenarios at the LHC.  Comparing limits, we find that current LHC results only exclude regions in parameter space which were already disfavored by precision electroweak data.  

\end{abstract}

\maketitle

\section{Introduction}

Vital clues to the nature of electroweak symmetry breaking have emerged from the LHC.  The bulk of the standard model (SM) Higgs mass region has been excluded at 95\% CL \cite{ATLAS, CMS}, leaving a narrow window $123\text{ GeV} < m_h < 128$ GeV in which there is a modest excess of events consistent with $m_h \simeq 125$ GeV.
As is well-known, 
such a mass
 can be accommodated within the minimal supersymmetric standard model (MSSM) but this requires  large A-terms or very heavy scalars,
 which tend to destabilize the electroweak hierarchy and undermine the original naturalness motivation of supersymmetry (SUSY)  \cite{Djouadi, Draper, Ruderman}.  
 Post LEP, however, a variety of strategies were devised in order to lift the Higgs mass. In these models the Higgs quartic coupling is boosted: either at tree level, via non-decoupling F-terms \cite{NomuraHall, Murayama,Ellwanger} and D-terms \cite{Tait,WackerPierce}, or radiatively, via loops of additional matter \cite{Graham,Martin}.  Already, a number of groups have redeployed these model building tactics 
in light of the recent LHC Higgs results 
\cite{Ruderman,Ellwanger125, Arvanitaki,Endo}.

The present work explores non-decoupling D-terms in gauge extensions of the MSSM. 
Our aim is to identify the prospects for observing this scenario at the  LHC in a maximally model independent way.
To begin, consider the MSSM augmented by an arbitrary flavor universal $U(1)_X$, which may be parameterized as a linear combination of  hypercharge $Y$, Peccei-Quinn number $PQ$, baryon number $B$, and lepton number $L$.  The Higgs must carry $X$ charge if the
corresponding D-terms are to contribute to the Higgs potential, so $X$ must have a component in $Y$ or $PQ$.  However, $PQ$ forbids an explicit $\mu$ term, so gauging $PQ$ requires a non-trivial modification to the Higgs sector which is highly model dependent.  To sidestep this complication we ignore $PQ$ and study the otherwise general space of $U(1)_X$ theories consistent with a $\mu$ term,
 \bea
 X &=& Y + p B - q L,
 \label{YBL}
 \eea
where the normalization of $X$ relative to $Y$ has been absorbed into the sign and magnitude of the gauge coupling, $g_X$. 
 We impose no further theoretical constraints, but will comment later on anomalies, naturalness, and perturbative gauge coupling unification. 
 As we will see, the ultraviolet dynamics, e.g.~the precise mechanism of gauge symmetry and SUSY breaking, will be largely irrelevant to our analysis.

 We constrain $U(1)_X$ with experimental data from resonance searches, precision electroweak measurements, and Higgs results.
 Remarkably, non-trivial limits can be derived without exact knowledge of seemingly essential parameters like $g_X$, $p$, and $q$.
This is possible because $g_X$ is bounded {\it from below} by the mass of the Higgs while  the couplings of the $X$ boson to quarks, leptons, and the Higgs  are {\it non-zero for all} values of $p$ and $q$.  As a result, for a fixed value of the $X$ boson mass, $m_X$, the theory predicts an irreducible rate, $\sBR$, for the process $pp\rightarrow X \rightarrow \ell\ell$ (relevant to direct searches) and an irreducible coupling of $X$ to the Higgs and leptons (relevant to precision electroweak data).

\begin{figure}[t]
\centering
\hspace*{-0.75cm}
\includegraphics[scale=1.45]{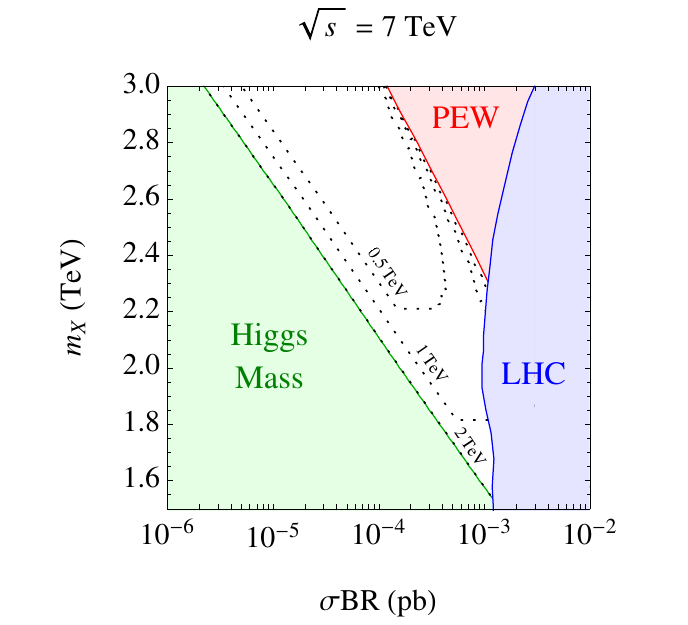} \\
\caption{\footnotesize{Litmus test: parameter space excluded by precision electroweak measurements (red),  Higgs mass limits (green), and LHC resonance searches (blue) at $\sqrt{s}=7$ TeV.  For $\sBR$ too large, $g_X > \gxmax$ yielding tension with precision electroweak and LHC constraints; for $\sBR$ too small, $g_X < \gxmin$ yielding tension with $m_h\simeq 125$ GeV subject to the stop mass, shown here for $m_{\tilde t} = 0.5\text{ TeV, } 1\text{ TeV, } 2 \text{ TeV}$.  
See the text in \Sec{sec:result} for details.  }}
\label{fig:main7}
\end{figure}

Combining limits,  we derive a maximal allowed region in the space of observables, $(\sBR,m_X)$, outside of which is either unnatural or in conflict with experimental bounds, as shown in \Figs{fig:main7}{fig:main14}.  If non-decoupling D-terms indeed play a role in boosting the Higgs mass, then the $X$ boson can only be observed within this allowed region---a `litmus test' for this general class of theories.
Furthermore, we find that for natural SUSY, i.e.~$m_{\tilde t} \lesssim 500$ GeV, resonance searches from the LHC \cite{LHCres} are not yet competitive with existing precision electroweak constraints. 

In \Sec{sec:setup} we define our basic setup.  Applying the constraints of gauge symmetry and SUSY, we  derive a general expression for the Higgs potential arising from non-decoupling D-terms.    Afterwards, in \Sec{sec:result} we compute the  Higgs mass and the couplings of the MSSM fields to the $X$ boson.  We then impose experimental limits and suggest a simple litmus test for non-decoupling D-terms. We conclude in \Sec{sec:conclusions}.


\section{Setup}

\label{sec:setup}


We are interested in all $U(1)_X$ extensions of the MSSM consistent with  a gauge invariant $\mu$ term.  
 Mirroring \cite{Appelquist, Villadoro}, we go to a convenient basis in which
the charge parameters, $g_X$, $p$, and $q$, absorb all of the effects of kinetic and mass mixing between the $U(1)_X$ and $U(1)_Y$ gauge bosons above the electroweak scale.  Thus, mixing only occurs after electroweak symmetry breaking, and the resulting effects are proportional to the Higgs vacuum expectation value (VEV).  Of course,  kinetic mixing is  continually induced by running, so  this choice of basis is renormalization scale dependent.  However, this subtlety is largely irrelevant to our analysis, which involves experimental limits in a relatively narrow window of energies around the weak scale.   The advantage of this low energy parameterization is that it is very general and covers  popular gauge extensions like $U(1)_B$, $U(1)_L$, $U(1)_{B-L}$, $U(1)_\chi$, and $U(1)_{3R}$.  Furthermore, it is defined by a handful of parameters: $m_X$, $g_X$, $p$, and $q$.

\begin{figure}[t]
\centering
\hspace*{-0.75cm}
\includegraphics[scale=1.45]{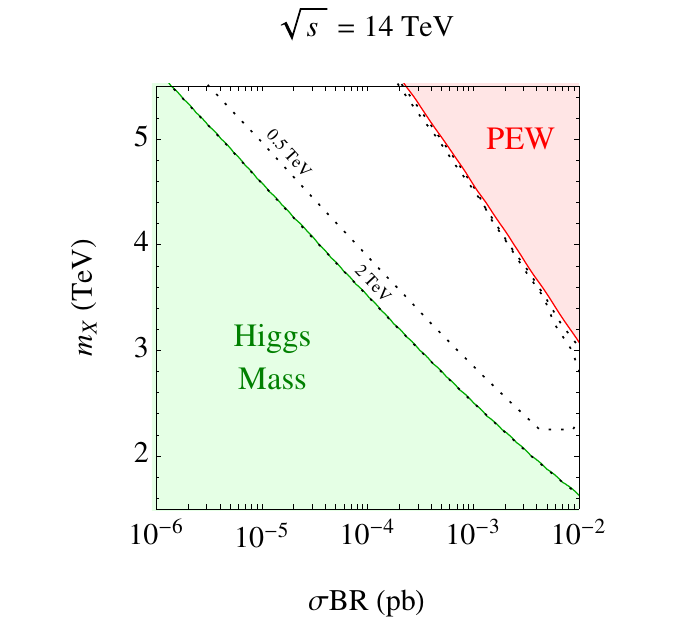} \\
\caption{\footnotesize{Same as \Fig{fig:main7} except with $\sqrt{s} = 14$ TeV, and stop mass contours $m_{\tilde t} = 0.5\text{ TeV, }2\text{ TeV}$.}}
\label{fig:main14}
\end{figure}

Next, let us consider the issue of anomalies.
If $p=q$, then according to  \Eq{YBL} $X$ is a linear combination of the $Y$ and $B-L$, which is anomaly free if one includes a flavor triplet of right-handed neutrinos.  If $p\neq q$ then the associated $B+L$ anomalies can be similarly cancelled by new particles.   In general, these `anomalons' can be quite heavy, in which case they can be ignored for  our analysis.

\begin{figure}[t]
\centering
\includegraphics[scale=1.35]{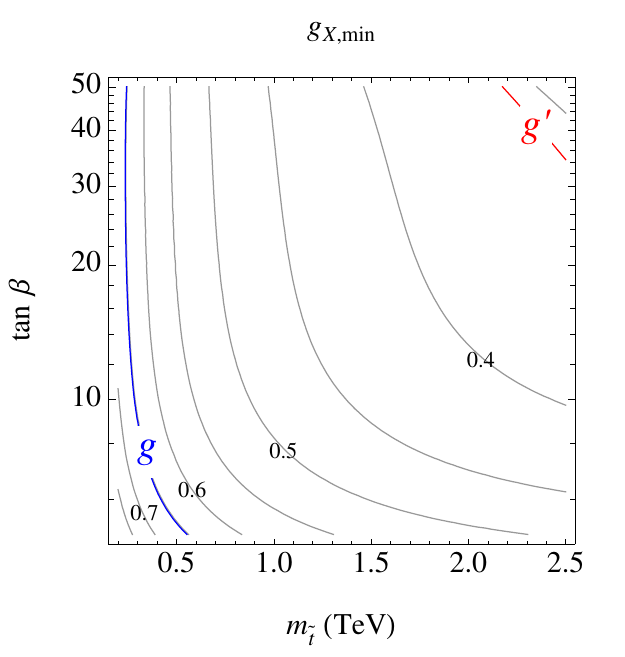} \\
\caption{\footnotesize{Contours of $\gxmin$ which set the lower bound on $g_X$ required to raise the Higgs mass to $m_h \simeq 125$ GeV.  Values equal to the SM gauge couplings are highlighted.  
}}
\label{fig:gxmin}
\end{figure}
\begin{figure}[t]
\centering
\includegraphics[scale=1.35]{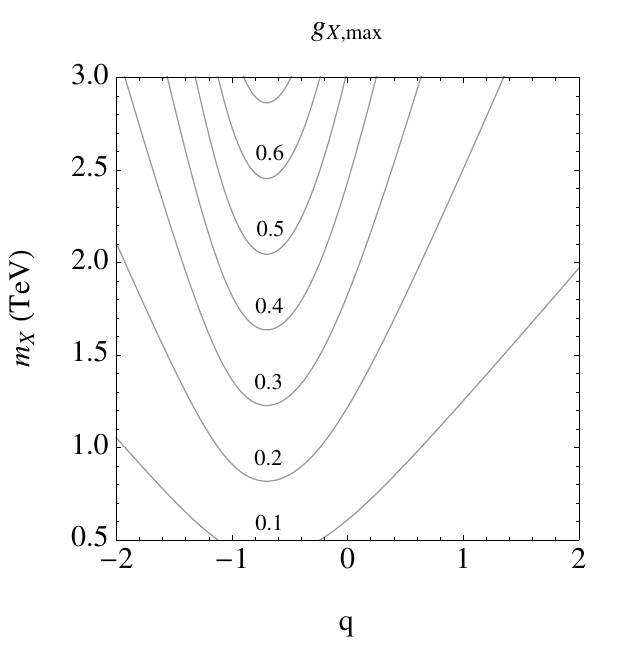} \\
\caption{\footnotesize{Contours of $\gxmax$ which set the upper bound on $g_X$ dictated by precision electroweak constraints.  These limits depend primarily on the couplings of $X$ to leptons, which are set by the $q$ parameter.
}}
\label{fig:gxmax}
\end{figure}

We now examine the non-decoupling D-terms of $U(1)_X$ and their contribution to the Higgs potential.  As we will see, these contributions are highly constrained by gauge symmetry and SUSY.
To begin, consider a massive vector superfield composed of component fields
\bea
 \{C, \chi, X, \lambda, D\},
\eea
where $X$, $\lambda$, and $D$ are the gauge field, gaugino, and auxiliary field, and
 $C$ and $\chi$ are the `longitudinal' modes eaten during the super-Higgs mechanism.
Under SUSY transformations, 
\bea
C& \rightarrow& C + i (\xi \chi - \bar\xi \bar\chi) \\
D &\rightarrow& D + \partial_\mu (-\xi \sigma^\mu \bar\lambda + \lambda\sigma^\mu \bar\xi).
\label{SUSYtransform}
\eea
\Eq{SUSYtransform} implies that $m\, C - D  $ is a SUSY invariant on the equations of motion, $i \sigma^\mu \partial_\mu \bar \lambda = m \chi$, where $m= m_C = m_\lambda = m_X$ is the mass of the vector superfield.     

On the other hand, the auxiliary field $D$ can be re-expressed in terms of dynamically propagating fields by substituting the equations of motion.  Since $m\,C-D$ is a SUSY invariant, this implies that
\bea
D =    m\,C +D_{\rm IR} + D_{\rm UV}+ \OO(C^2),
\label{eq:Ddef}
\eea
where  $D_{\rm IR}$ and $D_{\rm UV}$ label contributions from the (light) MSSM fields and the (heavy) $U(1)_X$ breaking fields, respectively, with all $C$ dependence shown explicitly.  The structure of  \Eq{eq:Ddef} ensures that both the right and left hand sides transform the same under SUSY transformations.
In the normalization of \Eq{YBL}, $H_{u,d}$ has charge $\pm 1/2$ under $U(1)_X$, which implies
\bea
D_{\rm IR} &=& \frac{g_X}{2} (|H_u|^2 -|H_d|^2 + \ldots).
\label{eq:Dhiggs}
\eea
The effective potential for $C$ and the MSSM scalars is obtained by setting all other fields to their VEVs, yielding
\bea
V &=&  \frac{1}{2}D^2 + \frac{1}{2} \tilde m^2  C^2 + \tilde t C.
\label{Dmass}
\eea
The first term is the usual SUSY D-term contribution, while the second and third terms arise from soft SUSY breaking effects such as non-zero F-terms.  Here we have  dropped terms $\OO(C^3)$ and higher because they are unimportant for the Higgs quartic. 
Note that the spurions $\tilde m$ and $\tilde t$ depend implicitly on the VEVs of $U(1)_X$ breaking sector fields.  

In the SUSY limit, $\tilde m = \tilde t  = 0$ and integrating out $C$ eliminates all $D_{\rm IR}$ dependence in the potential---no Higgs quartic is induced, as expected.  
If, on the other hand, $\tilde t \neq 0$, then $C$ and $D_{\rm UV}$ will typically acquire messenger scale VEVs, yielding a huge tree-level contribution to $m_{H_u}$ and $m_{H_d}$ through a term linear in $D_{\rm IR}$.  To avoid a destabilization of the electroweak scale, one usually assumes some ultraviolet symmetry, e.g.~messenger parity, which ensures $\tilde t = 0$ and vanishing VEVs for $C$ and $D_{\rm UV}$.  We assume this to be the case here, in which case there is no D-term SUSY breaking. 

On the other hand,  SUSY breaking typically enters through $\tilde m\neq 0$, whose effects can be characterized by a simple SUSY spurion analysis.  Let us model $\tilde m$ by an ultraviolet superfield spurion for F-term breaking, $\theta^2 F$.  This spurion can effect the scalar sector in two ways:  through the indirect shifts of scalar component VEVs, or through the direct couplings of $\theta^2 F$ to superfields.   In the former, the masses of $C$ and $X$ may vary, but they do so together, and the states remain degenerate.  In the latter, only certain couplings are permitted between $\theta^2 F$ and the vector superfield components.  Simple $\theta$ and $\bar\theta$ counting shows that $X$ and $D$ cannot couple directly to $\theta^2F$, while $C$ can.   Hence, $C$ is split in mass from the remainder of the gauge multiplet by F-term SUSY breaking.

Putting this all together, we rewrite \Eq{Dmass} as
\bea
V &=&  \frac{1}{2}\left( m_X C + D_{\rm IR}\right)^2 + \frac{1}{2}  (m_C^2-m_X^2)  C^2,
\eea
where the coefficient of the second term is fixed so that $m_C$ is the physical mass of $C$.  Note that the prefactor for $C$ in the first term is $m_X$---this can be verified by explicit computation, and is a direct consequence of the fact that $X$ and $D$ cannot couple directly to $\theta^2 F$.  
Integrating out $C$ yields our final answer for the effective D-term contribution to the Higgs potential
\bea
V&=& \frac{1}{2}\varepsilon D_{\rm IR}^2 \\
\varepsilon &=& 1 - m_X^2/m_C^2
\label{varepsilon},
\eea
which is a generalization of the specific examples in \cite{Tait,WackerPierce}.
In the SUSY limit, $m_C = m_X$ and the D-term contribution vanishes as expected. A positive contribution to the Higgs mass requires positive $\varepsilon$, which in turn requires that $m_C > m_X$.
Importantly, $0\leq \varepsilon < 1$ {\it independent of the ultraviolet completion}, which will be crucial later on when we derive model independent bounds.

\begin{figure}[t]
\centering
\includegraphics[scale=1.35]{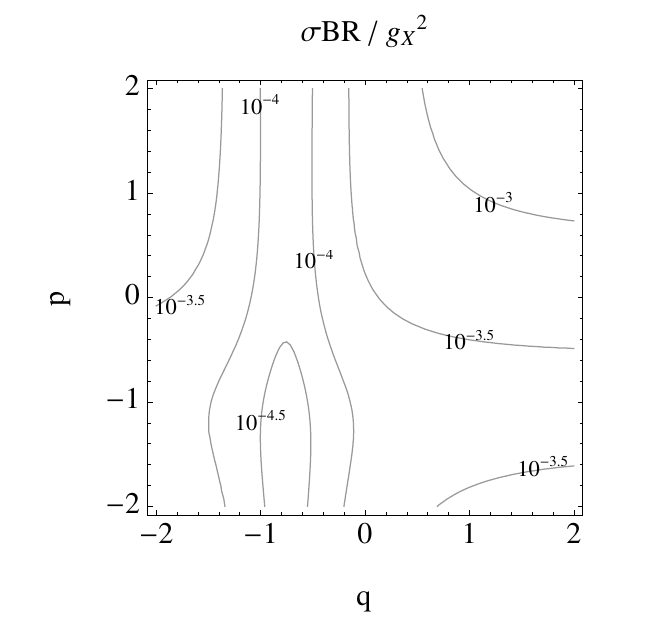} \\
\caption{\footnotesize{Contours of $\sBR/g_X^2$ in pb for $m_X = 3$ TeV and $\sqrt{s} = 7$ TeV.  Irrespective of the $U(1)_X$ charge parameters $p$ and $q$, the rate is {\it always} non-zero.}}
\label{fig:sigmaBR7}
\end{figure}

\section{Results}

\label{sec:result}

\subsection{Experimental Constraints}

\label{sec:exp}

In this section we analyze the experimental constraints on general $U(1)_X$ extensions of the MSSM.  The relevant bounds come from the mass of the Higgs boson, precision electroweak measurements, and direct limits from the LHC.

 $\bullet$ {\it Higgs  Boson Mass.}   Recent results from the LHC indicate hints of a SM-like Higgs boson at around $m_h \simeq 125$ GeV.  Taken at face value, this imposes a stringent constraint on theories of $U(1)_X$ D-terms.
In particular, combining \Eq{eq:Dhiggs} with \Eq{varepsilon} yields the mass of the Higgs boson
\bea
m_h^2 &=& m_Z^2 \cos^2 2 \beta \left(1+  \frac{\varepsilon g_X^2}{g'^2 + g^2}\right) + \delta m_h^2,
\label{mh}
\eea
where $0 \leq \varepsilon <1$ independent of the ultraviolet completion. Here $\delta m_h^2$ denotes the usual radiative contributions to the Higgs mass in the MSSM,
\bea
\delta m_h^2 &=& \frac{3 m_t^4}{4 \pi^2 v^2}\left(\log \frac{m_{\tilde t}^2}{m_t^2} + \frac{X_t^2}{m_{\tilde t}^2}\left(1-\frac{X_t^2}{12m_{\tilde t}^2}\right)
\right) ,
\eea
where $m_{\tilde t} = (m_{\tilde t_1} m_{\tilde t_2})^{1/2}$ and $X_t = A_t - \mu \cot \beta$.  In our actual analysis we employ the  analytic expressions from \cite{Carena} for the Higgs mass, which include two-loop leading log corrections.

\begin{figure}[t]
\centering
\includegraphics[scale=1.35]{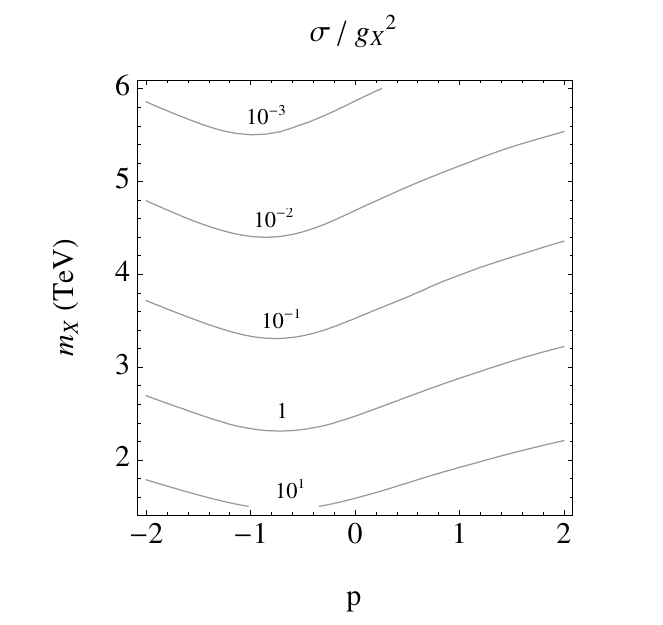} \\
\caption{\footnotesize{Contours of $\sigma/g_X^2$ in pb for $\sqrt{s} = 14$ TeV.}}
\label{fig:sigma14}
\end{figure}

\begin{figure}[t]
\centering
\includegraphics[scale=1.35]{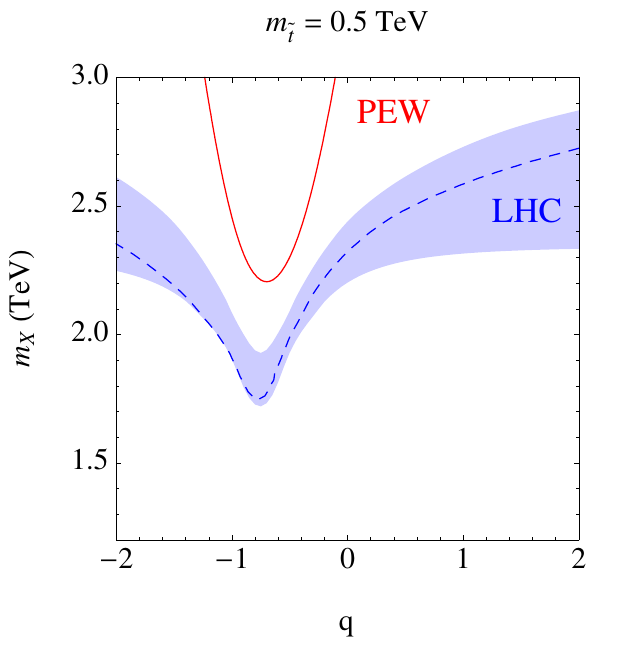} \\
\caption{\footnotesize{Limits from precision electroweak measurements (red solid), and LHC resonance searches for $p=q$ (blue dashed) and $p\neq q$ free (blue shaded), with $m_{\tilde t}= 0.5$ TeV, corresponding to $g_X > \gxmin = 0.54$. Direct searches only exclude regions already disfavored by precision electroweak constraints.}}
\label{fig:combo1}
\end{figure}

\begin{figure}[t]
\centering
\includegraphics[scale=1.35]{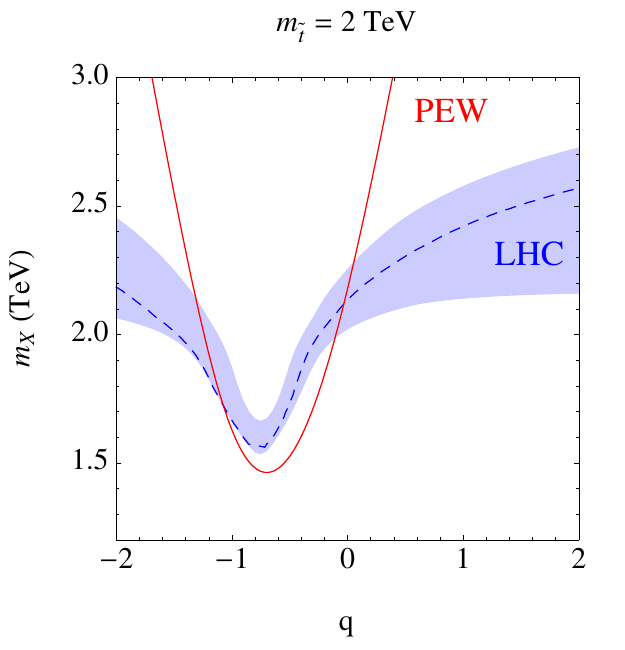} \\
\caption{\footnotesize{Same as \Fig{fig:combo1}, but for $m_{\tilde t} = 2$ TeV, corresponding to $g_X > \gxmin = 0.36$.}}
\label{fig:combo2}
\end{figure}

To simplify the parameter space, we take $A_t = 0$ and $\mu = 200$ GeV.  Our results will be indicative of theories which have small A-terms, such as gauge mediated SUSY breaking.
 For  a given value of $\tan \beta$ and $m_{\tilde t}$, the Higgs mass correction $\delta m_h^2$ is then fixed.   Using \Eq{mh}  and $0 \leq \varepsilon < 1$, we find that $g_X$ is bounded from below in order to accommodate $m_h \simeq 125$ GeV:
 \bea
  g_X >\gxmin ,
  \label{eq:gxmin}
  \eea
  where $\gxmin$ is a function of $(m_{\tilde t},\tan \beta)$ shown in  \Fig{fig:gxmin}.  For comparison, this figure includes contours of the SM electroweak gauge couplings, $g'$ and $g$.    At high  $\tan \beta$, $U(1)_X$ is most effective at lifting the Higgs mass, so the stop masses can be the smallest.
Note that in certain ultraviolet completions, $\varepsilon$ can be quite small, in which case $\gxmin$ and thus $g_X$ will be much larger than the SM gauge couplings.

Lastly, let us  comment briefly   on the issue of fine tuning.  In \Sec{sec:setup} we showed that non-decoupling D-terms require the scalar $C$ to be split from the $X$ boson at tree level.  As a consequence, the low energy Higgs quartic coupling behaves like a hard breaking of SUSY and loops involving the components of the vector supermultiplet generate a quadratic divergence which is cut off by $m_X$.  Since the Higgs fields are charged under $U(1)_X$, these radiative corrections contribute to the Higgs soft masses at one loop and can destabilize the electroweak hierarchy.  In particular,
\bea
\delta m_{H_{u,d}}^2 &=& \frac{g_X^2 }{64\pi^2}  m_{X}^2  \log \left(\frac{m_{X}^6  m_C^2 }{m_{\lambda}^8}\right),
\label{eq:tuning}
\eea
which applies to R-symmetric limit \cite{Sudano,ZoKo}.  As required, when the components of the supermultiplet become degenerate, these corrections vanish.   Due to the loop factor in  \Eq{eq:tuning} and the relative smallness of $g_X$ required to lift the Higgs mass in \Fig{fig:gxmin}, $m_X$ can be quite large---even beyond LHC reach---without introducing  fine-tuning more severe than $\sim 10\%$.

$\bullet$ {\it Precision Electroweak \& Direct Limits.}  Contributions to precision electroweak observables arise from two sources: mixing between the $X$ and $Z$ bosons, and  couplings between the $X$ boson and leptons.  The former is always generated by electroweak symmetry breaking since the Higgs is, by construction, charged under $U(1)_X$.  Meanwhile, the latter is also always present, since $X$ has an irreducible coupling to leptons.  Concretely, since $H_{u,d}$ has charge $\pm1/2$, this implies that the composite operators $QU^c$, $QD^c$, and $LE^c$ have charge $-1/2$, $+1/2$, and $+1/2$, respectively.  As a result, $X$ has an irreducible coupling to both leptons and quarks.  The branching ratio to a single lepton flavor is:
\bea
\text{BR}(X\rightarrow \ell \ell)\simeq \frac{5 + 12 q + 8 q^2}{66 + 24 p + 24 p^2 + 72 q + 54 q^2},
\label{eq:BRs}
\eea
where we have ignored kinematic factors and have assumed that the full MSSM field content can be produced in the decays of the $X$ boson.  
This is a conservative choice because decoupling MSSM fields always increases $\text{BR}(X\rightarrow \ell \ell)$, yielding more stringent constraints.  For example, if $X$ decays to the first and second generation squarks are kinematically forbidden, then $\text{BR}(X\rightarrow \ell \ell)$ will increase at most by a factor of $\sim 1.2$.  Using \Eq{eq:BRs}, we see that the leptonic branching ratio never vanishes for any finite values of $p$ and $q$, and is strictly bounded from above at $\sim 15$\%.

Applying the methods of \cite{Strumia}, we performed a precision electroweak fit on the theory parameters, $g_X/m_X$ and $q$.  For simplicity, we assumed a decoupling limit in which the lighter Higgs doublet drives the fit, so the Higgs sector is SM-like. 
As noted in \cite{Strumia}, the resulting constraints are dominated by the couplings of $X$ to leptons and the Higgs and are thus independent of $p$ to a very good approximation.   We have checked that our results match \cite{Villadoro}, which studies precision electroweak constraints on anomaly free $U(1)$ extensions.
To accommodate 95\% CL exclusion limits,  the gauge coupling is bounded from above by
 \bea
  g_X <\gxmax ,
  \label{eq:gxmax}
  \eea
where $\gxmax$ is a function of $(q, m_X)$ shown in  \Fig{fig:gxmax}.  Bounds are weakest near $q\simeq -0.7$ which is where the $Y$ and $L$ components of the $X$ charge destructively interfere in a way that decreases the effective coupling of the $X$ boson to leptons.
  
Lastly, for LHC resonance searches we are interested in the rate of resonant production, $\sBR$ for the process $pp \rightarrow X\rightarrow \ell \ell$.  
The leptonic branching ratios are given in \Eq{eq:BRs} as a function of $p$ and $q$, while the production cross-section of $X$ bosons from proton collisions can be  computed in terms of $p$ with {\tt MadGraph5}, including NNLO corrections from \cite{King}. 
Remarkably, $\sBR$ is non-zero for any value of $p$ and $q$, as shown in \Fig{fig:sigmaBR7},
which shows the rate normalized to $g_X^2$ for a sample parameter space point,  $m_X = 3$ TeV
at $\sqrt{s} = 7$ TeV. This crucially implies an irreducible rate for $pp \rightarrow X\rightarrow \ell \ell$, which we constrain with 5/fb results from the LHC \cite{LHCres}. For convenience, we also present the production cross-section normalized to $g_X^2$ in \Fig{fig:sigma14}.  By multiplying by BR($X\rightarrow \ell\ell$) from \Eq{eq:BRs} and $g_X^2$ which is bounded from \Figs{fig:gxmin}{fig:gxmax}, one can determine a simple estimate for the future LHC reach for $X$ bosons.  At 100/fb and $\sqrt{s}= 14$ TeV,  the LHC can reach as high as $m_X \sim 6$ TeV.


\begin{figure}[t]
\centering
\includegraphics[scale=1.35]{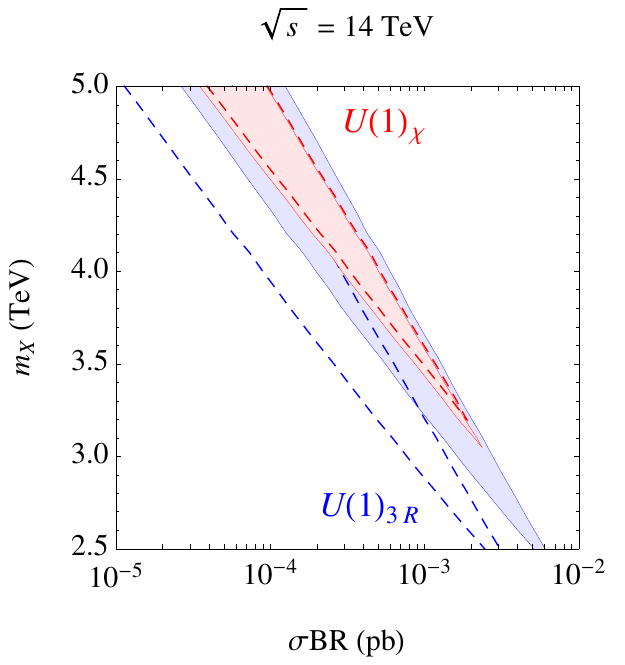} \\
\caption{\footnotesize{Allowed regions for $U(1)_\chi$ and $U(1)_{3R}$ for $p$ and $q$ fixed according to the exact GUT relations (solid shaded)  or fixed to their values when running down from a high scale (dashed). }}
\label{fig:GUT}
\end{figure}

\subsection{Litmus Tests}

The experimental constraints enumerated in \Sec{sec:exp} provide stringent and complementary limits on the allowed parameter space of $U(1)_X$ theories.  We can now combine these bounds in order to identify various `litmus tests' for non-decoupling D-terms.

To begin, consider \Figs{fig:combo1}{fig:combo2}, which depict experimentally excluded regions in the $(q,m_X)$ plane  for $m_{\tilde t} = 0.5\text{ TeV}, 2$ TeV, respectively.  The region below the solid red line is excluded by precision electroweak measurements.  This limit is to good approximation independent of $p$, which controls the coupling of $X$ to quarks.  The region below the blue dashed line is excluded by LHC resonance searches in the anomaly free case, i.e.~$p=q$.  Allowing $p\neq q$ to vary freely then floats the boundary of this exclusion within the blue shaded region. 

For stop masses in the natural window, $m_{\tilde t} \lesssim 500$ GeV, these plots imply that the LHC has not excluded any region of parameter space which was not already disfavored by precision electroweak limits.   Conversely, if natural SUSY employs non-decoupling D-terms, then the LHC should not yet have seen any signs of the $X$ boson.  Given precision electroweak measurements, $m_X \gtrsim 2.2$ TeV for natural SUSY.  For heavier stop masses, \Fig{fig:combo2} shows that the LHC has covered some but not very much new ground.

Let us now discuss \Figs{fig:main7}{fig:main14}.  At fixed values of the masses, $m_X$ and $m_{\tilde t}$, we can  scan over the charge parameters, $g_X$, $p$, and $q$, discarding any model points which are in conflict with precision electroweak and Higgs limits.
By this procedure, we obtain an `image' of the viable theory space  on the  observable space, $(\sBR, m_X)$.  Each dotted black contour in  \Figs{fig:main7}{fig:main14} depicts a maximal allowed region in $(\sBR, m_X)$ obtained via this scan for a given stop mass.    Any theory of natural SUSY which employs non-decoupling D-terms predicts an $X$ boson residing somewhere within the region corresponding to $m_{\tilde t} = 0.5$ TeV.   Since we have marginalized over $g_X$, $p$, and $q$, these exclusions are model independent.

The allowed regions in  \Figs{fig:main7}{fig:main14} are bounded at small and large $\sBR$ because $\gxmin < g_X < \gxmax$, where $\gxmin$ is a function of $(m_{\tilde t},\tan \beta)$
and $\gxmax$ is a function of $(q, m_X)$.  As described in \Sec{sec:exp}, the lower bound arises from the requirement that non-decoupling D-terms sufficiently lift the Higgs mass up to $m_h\simeq 125$ GeV, while the upper bound arises from precision electroweak constraints.
Since the production cross-section of $X$ bosons depends on $g_X$, one can translate this allowed window in $g_X$ into an allowed window in rate, $\sBR_{\rm min} < \sBR < \sBR_{\rm max}$.  

Because  \Figs{fig:main7}{fig:main14} were derived from a parameter scan, model points near the Higgs boundary limit versus those near the precision electroweak boundary limit correspond to different values of $p$ and $q$. This results in different precision electroweak constraints for different stop masses---an effect that is  amplified on the near flat direction in $m_{\tilde t}$ that traverses diagonally across the plot.


Note that the values of $\sBR_{\rm min}$ depicted in \Figs{fig:main7}{fig:main14} are   conservative---they coincide with the parameter choice  $\varepsilon = 1$ in \Eq{varepsilon}.  Because this corresponds to $m_C \rightarrow \infty$, this choice is rather unphysical.  
In general, $\varepsilon < 1$, in which case $\sBR_{\rm min}$ will be substantially larger and the allowed region will shrink.

Also, at a fixed value of $\sBR$, increasing $m_X$ makes precision electroweak bounds more severe, which is unintuitive from the point of view of decoupling.  However, this occurs because in order to keep $\sBR$ constant with increasing $m_X$, the coupling $g_X$ must increase even faster, inducing tension with precision electroweak measurements.



Alternatively, we can fix $p$ and $q$ rather than marginalize with respect to them.  GUT relations provide a natural choice for the values of $p$ and $q$:
\bea
U(1)_\chi &:& p=q = -5/4 \\
U(1)_{3R} &:& p=q =-1/2. 
\label{eq:exactGUT}
\eea
However,  running from high scales can induce kinetic mixing which offsets $p$ and $q$, which are intrinsically low energy parameters.  For $m_{\rm GUT} \simeq 2\times 10^{16}$ GeV, this can shift $p=q$ up to about $-1.2$ for $U(1)_\chi$ and down to about $-0.8$ for $U(1)_{3R}$, although the precise numbers depend on the GUT scale and matter content \cite{Villadoro}.   Because GUT values may be preferred from a top down viewpoint, we present  the allowed regions for these theories at $\sqrt{s}= 14$ TeV  in \Fig{fig:GUT}, depicted as the colored wedges.  As before, lower values of $\sBR$ are excluded by the Higgs mass results (where here we have fixed $m_{\tilde t}=0.5$ TeV) while higher values of $\sBR$ are excluded by precision electroweak constraints.
Theories corresponding to the exact GUT values for $p=q$ in \Eq{eq:exactGUT} are depicted by solid lines, while the dashed lines depict values of $p=q$ including running from high scales.  For both $U(1)_\chi$ and $U(1)_{3R}$, a narrow allowed region is prescribed, outside of which is either unnatural or experimentally excluded.

\section{Conclusions}

\label{sec:conclusions}

In this paper we have analyzed a broad class of $U(1)_X$ extensions of the MSSM in which $m_h \simeq 125$ GeV is accommodated by non-decoupling D-terms.  We have assumed that $U(1)_X$ is flavor universal and allows a gauge invariant $\mu$ term, but impose no additional theoretical constraints.  

Our main result is a simple litmus test for this class of theories at the LHC---if non-decoupling D-terms are instrumental in lifting the Higgs mass, then experimental constraints imply that an $X$ boson can only be observed in the  allowed region depicted in \Figs{fig:main7}{fig:main14}.  
Crucially, for natural SUSY this region is bounded from below in $\sBR$ for $pp\rightarrow X \rightarrow \ell\ell$, so we should expect an irreducible level of $X$ boson production at the LHC.   Our check is very model independent,  since our input constraints have been marginalized over all charge assignments for $U(1)_X$.  Furthermore, general arguments from SUSY and gauge invariance dictate the very particular form for non-decoupling D-terms shown \Eq{varepsilon}, so our results are also independent of the ultraviolet details of $U(1)_X$ breaking.   We have also presented an analogous litmus test which can be applied for the specific GUT inspired models described in \Fig{fig:GUT}.
\begin{center}
{\bf \small Acknowledgments}
\end{center}

C.~C.~and H.~R.~are supported by the Director, Office of Science, Office of High Energy and Nuclear Physics, of the US Department of Energy under Contract DE-AC02-05CH11231, and by the National Science Foundation under grant PHY-0855653.   C.~C.~would like to thank Josh Ruderman for useful comments.

\end{document}